\documentstyle[12pt]{article}

\input epsf

\ifx\epsffile\undefined
\newlength{\epsfysize}
\def\epsffile#1#2#3#4]#5{}
\else
\fi

\def\roughly#1{\raise.3ex\hbox{$#1$\kern-.75em\lower1ex\hbox{$\sim$}}}
\def\o{\over}
\def\dr{\mbox{$\overline{\it DR}$}~}
\def\ms{\mbox{$\overline{\it MS}$}~}

\begin{document}
\begin{titlepage}
\begin{center}
January, 1995 \hfill       JHU-TIPAC-95001\\
\hfill hep-ph/9501277\\
\vskip .7 in
{\large \bf Precision Corrections to Supersymmetric Unification}
\vskip .3 in
           \vskip 0.5 cm
      {\bf\centerline{ Jonathan Bagger\qquad Konstantin Matchev} \vskip.4cm
          \centerline{ and Damien Pierce  } }     \vskip.4cm
      {\it Department of Physics and Astronomy\\
          The Johns Hopkins University\\
         Baltimore, Maryland\ \ 21218\\}

\end{center}
\vskip 0.4 in
\begin{abstract}
We compute the full set of weak-scale gauge and Yukawa threshold
corrections in the minimal supersymmetric standard model, including
all finite (non-logarithmic) corrections, which we show to be
important.  We use
our results to examine the effects of unification-scale threshold
corrections in the minimal and missing-doublet SU(5) models.  We work
in the context of a unified mass spectrum, with scalar mass $M_0$ and
gaugino mass $M_{1/2}$, and find that in minimal SU(5)
with squark masses less than one TeV, successful
gauge and Yukawa coupling unification requires $M_{1/2}\ll M_0$ and
$M_0\simeq1$ TeV. In contrast, we find that the
missing-doublet model permits gauge and Yukawa unification for
a wide range of supersymmetric masses.
\end{abstract}
\end{titlepage}

\renewcommand{\thepage}{\arabic{page}}
\setcounter{page}{1}

\section{Introduction}

With the advent of precision measurements at LEP and the observation
that the gauge couplings unify in the minimal supersymmetric standard model
\cite{unified}, there has been a resurgence of interest in supersymmetric
grand unified theories.  In most analyses the values of the electromagnetic
coupling, $\alpha_{\rm EM}$, and the standard-model weak mixing angle, $
s^2_{\rm SM} \equiv \sin^2\theta_{\rm SM}(M_Z)$, are taken from experiment,
and converted into supersymmetric \dr parameters, $\hat\alpha$ and $\hat s^2
\equiv \sin^2\hat\theta(M_Z)$, using the leading-logarithmic contribution
to the supersymmetric
threshold corrections.  The \dr parameters are then used to determine the \ms
strong coupling constant at the $Z$-scale, $\alpha_s(M_Z)$, ignoring
all unification-scale threshold corrections.
Alternatively, the measured value of
$\alpha_s(M_Z)$ is used to constrain the unification-scale parameter space.

In this letter we will take a closer look at this procedure, and report
on the results of a complete next-to-leading-order analysis of supersymmetric
unification.  Our approach is new in two respects.  First, we include
all finite corrections.  This implies that our weak mixing angle $\hat s^2$
is related to $s^2_{\rm SM}$ as follows,
\begin{equation}
\hat s^2\ =\ s^2_{\rm SM}
\ + \ \rm leading~log~+~finite,
\label{sintheta}
\end{equation}
where the leading logarithms are of the form $\log(M_{\rm SUSY}/M_Z)$
and the finite corrections are of order $M_Z^2/M_{\rm SUSY}^2$.
If all the superpartner masses are heavier than a few times $M_Z$,
the finite corrections are negligible, in accord with the decoupling
theorem.  However, realistic supersymmetric models typically have light
particles with masses near $M_Z$, so the finite corrections can be
significant.

Second, we do not use a combined-fit value of $s^2_{\rm SM}$ to
compute $\hat s^2$ because the finite term in eq.~(\ref{sintheta})
is different for each observable. Therefore, in our analysis, we calculate
$\hat s^2$ directly
from a given set of inputs, namely, the Fermi constant $G_F$, the $Z$-boson
mass $M_Z$, the electromagnetic coupling $\alpha_{EM}$, the top-quark mass
$m_t$, and the parameters that describe the supersymmetric model.
(See also ref.~\cite{Warsaw}.)

We emphasize that a precise evaluation of $\hat s^2$ is essential because
the renormalization group equations (RGE's) can naturally amplify small
corrections.  For example, the one-loop renormalization group equations,
$d\hat g_i/dt = \beta_i\, \hat g_i^3/16 \pi^2$,
together with gauge coupling unification, imply
\begin{equation}
{\beta_2-\beta_3\o \hat g_1^2(\mu)}\ +\ {\beta_3-\beta_1\o \hat g_2^2(\mu)}\ +
\ {\beta_1-\beta_2\o \hat g_3^2(\mu)} \ =\ 0\ ,
\end{equation}
at any scale $\mu$.  Solving for $\alpha_s(M_Z)$ and varying the inputs
$\hat\alpha$ and $\hat s^2$, we find
\begin{equation}
{\delta\alpha_s\o\alpha_s(M_Z)} \ \simeq
\ \,{\delta\hat\alpha\o\hat\alpha} \ -
\ 7.5 \ {\delta\hat s^2\o\hat s^2} \ .
\label{delta_alpha}
\end{equation}
This shows that a 1\% error in the determination of $\hat s^2$
leads to an error in the evaluation of $\alpha_s(M_Z)$
of 7.5\%.

In what follows, we will briefly describe our calculation of $\hat\alpha$
and $\hat s^2$.  We will then examine our prediction for $\alpha_s(M_Z)$
in the supersymmetric parameter space.  We will find the unification-scale
threshold corrections that are necessary for gauge coupling
unification with a given
supersymmetric spectrum, and compare our results with the threshold
corrections that arise in the minimal and missing-doublet SU(5) models.
We will close with a similar analysis under the additional assumption of
bottom-tau Yukawa unification.

\section{Gauge Coupling Unification}

We start by sketching our calculation of the \dr gauge parameters
$\hat g_1(M_Z)$ and  $\hat g_2(M_Z)$.  As discussed above, we take as inputs
the Fermi constant $G_F = 1.16639 \times 10^{-5}$ GeV$^{-2}$,
the $Z$-boson mass $M_Z = 91.187$ GeV,
the electromagnetic coupling $\alpha_{\rm EM} = 1/137.036$,
the top-quark mass $m_t$, and the supersymmetric parameters.  From these we
calculate the electromagnetic coupling $\hat\alpha$ and the weak mixing angle
$\hat s^2$ in the \dr renormalization scheme,
\begin{equation}
\hat\alpha\ =\ {\alpha_{\rm EM}\o1-\Delta\hat\alpha}
\ ,\qquad\qquad\qquad\qquad \hat{s}^2\hat{c}^2\ = \
{\pi \alpha_{\rm EM}\o\sqrt2 G_F M_Z^2(1-\Delta\hat{r})}\ ,
\end{equation}
where $\hat c^2 = \cos^2\hat\theta(M_Z)$,
\begin{equation}
\Delta\hat\alpha\ =\ 0.0685 \pm 0.0007
\ -\ {\alpha_{\rm EM}\o2\pi}\Biggl\{
-{7\o4}\log\left(M_W\o M_Z\right)
\ +\ {16\o9}\log\left(m_t\o M_Z\right)
\ +\ {1\o3}\log\left(m_{H^+}\o M_Z\right)
\label{eq:da}
\end{equation}
$$
\ +\ \sum_{i=1}^6{4\o9}\log\left(m_{\tilde u_i}\o M_Z\right)
\ +\ \sum_{i=1}^6{1\o9}\log\left(m_{\tilde d_i}\o M_Z\right)
\ +\ \sum_{i=1}^3{1\o3}\log\left(m_{\tilde e_i}\o M_Z\right)
+ \sum_{i=1}^2{4\o3}\log\left(m_{\chi_i^+}\o M_Z\right)\Biggr\}\ ,
$$
and \cite{Degrassi}
\begin{equation}
\Delta\hat r\ =\ \Delta\hat\alpha\ +\ {\hat \Pi_W(0)\o M_W^2}
\ -\  {\hat \Pi_Z(M_Z)\o M_Z^2}\ +\ \hbox{vertex} \ + \ \hbox{box}
\ .  \label{eq:dr}
\end{equation}
Equation (\ref{eq:da}) includes the light quark contribution extracted from
experimental data \cite{Jegerlehner}, together with the leptonic
contribution. It also contains the logarithms of
the $W$-boson, top-quark, charged-Higgs, squark, slepton, and chargino masses.
In eq.~(\ref{eq:dr}), the $\hat\Pi$ denote the real and transverse parts of
the gauge boson self-energies, evaluated in the \dr scheme.  Equation
(\ref{eq:dr}) also includes the vertex and box contributions that renormalize
the Fermi constant, as well as the leading higher-order $m_t^4$ and QCD
standard-model corrections given in ref.~\cite{FKS}.
(For a complete description of our calculation see ref. \cite{BMPZ}.
Note that in this letter, all hatted objects are \dr quantities, and
all masses are pole masses.)

{}From these results we find the weak gauge couplings $\hat g_1(M_Z)$ and
$\hat g_2(M_Z)$ using the \dr relations
\begin{equation}
\hat g_1(M_Z)\ =\ \sqrt{5\o3}\,{\hat e\o\hat c}\ ,\qquad\quad
\hat g_2(M_Z)\ =\ {\hat e\o\hat s}\ ,
\end{equation}
where $\hat\alpha = \hat e^2/4\pi$.
These couplings serve as the weak-scale boundary conditions for the two-loop
supersymmetric RGE's.  They implicitly determine the
unification
scale $M_{\rm GUT}$ through the condition $\hat g_1(M_{\rm GUT}) = \hat
g_2(M_{\rm GUT}) \equiv g_{\rm GUT}$.

We fix the strong coupling $\hat g_3(M_{\rm GUT})$ at the scale
$M_{\rm GUT}$ by the unification condition
\begin{equation}
\hat g_3(M_{\rm GUT}) = g_{\rm GUT}(1 + \epsilon_g)\ ,
\label{ggut}
\end{equation}
where $\epsilon_g$ parametrizes the model-dependent unification-scale
gauge threshold correction.  We then run $\hat g_3(M_{\rm GUT})$
back to the $Z$-scale to find the \ms coupling
\begin{equation}
\alpha_s(M_Z)\ =\ \left({\hat g_3^2(M_Z)\o
4\pi}\right)\,(1-\Delta\alpha_s)^{-1}\ ,
\end{equation}
where $\Delta\alpha_s$ denotes the weak-scale threshold correction
\begin{equation}
\Delta\alpha_s\ =\ {\hat g_3^2(M_Z)\o8\pi^2}\Biggl(
-{1\o2} + {2\o3}\log\left(m_t\o M_Z\right)
+ \sum_{i=1}^{12}{1\o6}\log\left(m_{\tilde{q}_i}\o M_Z\right)
+ 2\log\left(m_{\tilde g}\o M_Z\right) \Biggr)\ .
\end{equation}

This procedure allows us to determine $\alpha_s(M_Z)$ for a fixed $\epsilon_g$
and a given supersymmetric spectrum.  In what follows, we will make the
additional assumption that the supersymmetric spectrum unifies as
well.  Therefore we also assume that the three gaugino masses unify
to a common value of $M_{1/2} \equiv \hat M_{1/2}(M_{\rm GUT})$, the scalar
masses unify to a common scalar mass $M_0 \equiv \hat M_0(M_{\rm GUT})$, and
the soft trilinear scalar coupling parameters unify to $A_0 \equiv \hat
A_0(M_{\rm GUT})$.

We evolve the parameters $M_{1/2},\ M_0$ and $A_0$ to the weak
scale using the two-loop supersymmetric RGE's \cite{two loop}.  We
require the parameters to
be such that electroweak symmetry is spontaneously broken, as naturally occurs
when the top-quark mass is large.  Then the Higgs bosons $H_1$ and $H_2$
acquire expectation values $v_1 \equiv \hat v_1(M_Z)$ and $v_2 \equiv \hat
v_2(M_Z)$; their ratio is denoted $\tan\beta \equiv \tan\hat\beta(M_Z) =
v_2/v_1$.

We extract the supersymmetric masses from the running \dr parameters and
$\tan\beta$ at the scale $M_Z$.  (We choose $\mu > 0$, where the
superpotential contains the term $+\mu\epsilon_{ij}H_1^i H_2^j,$ with
$\epsilon_{12} = +1$.)  We then substitute these masses into eqs.~(4) --
(6), and repeat the entire procedure to find a self-consistent
solution to the renormalization group equations.
Typically, the process converges after only a few iterations.  It allows us
to predict $\alpha_s(M_Z)$ consistently, including all finite corrections.

As a point of reference, we show in Fig.~1 our results for $\alpha_s(M_Z)$ in
the $M_0$, $M_{1/2}$ plane, in the absence of unification threshold
corrections, for
$m_t=170$ GeV, $\tan\beta=2$, and $A_0=0$.  Comparing with the experimental
value $\alpha_s(M_Z) = 0.117 \pm .005$ \cite{PDG}, we see that $\alpha_s(M_Z)$
is rather large.\footnote{The
experimental error in $\Delta\hat\alpha$ leads to a $\pm0.001$
uncertainty in $\alpha_s(M_Z)$.}
If, for naturalness, we require the squark masses
to be below 1 TeV, we obtain the lower bound
$$ \alpha_s(M_Z)\ >\ 0.126\quad\quad\quad ({\rm No\ unification\ thresholds},
\ m_{\tilde{q}}\le1{\rm\ TeV},\ m_t = 170 \rm\ GeV),$$
assuming the validity of perturbation theory, which we take to mean
that the \dr top-quark Yukawa coupling $\hat\lambda_t(M_{\rm GUT}) \le 3$.
If we tighten this condition to $\hat\lambda_t(M_{\rm GUT}) \le 1$, the
bound increases by 0.002.  For smaller $m_t$, $\alpha_s(M_Z)$
is smaller.  For example, if $m_t=160$ GeV, the limit reduces by 0.002.
These variations apply independently of the unification-scale thresholds.
Our numbers agree qualitatively with the results of ref.~\cite{Warsaw}.

\begin{figure}[tb]
\epsfysize=3.5in
\epsffile[30 420 600 735]{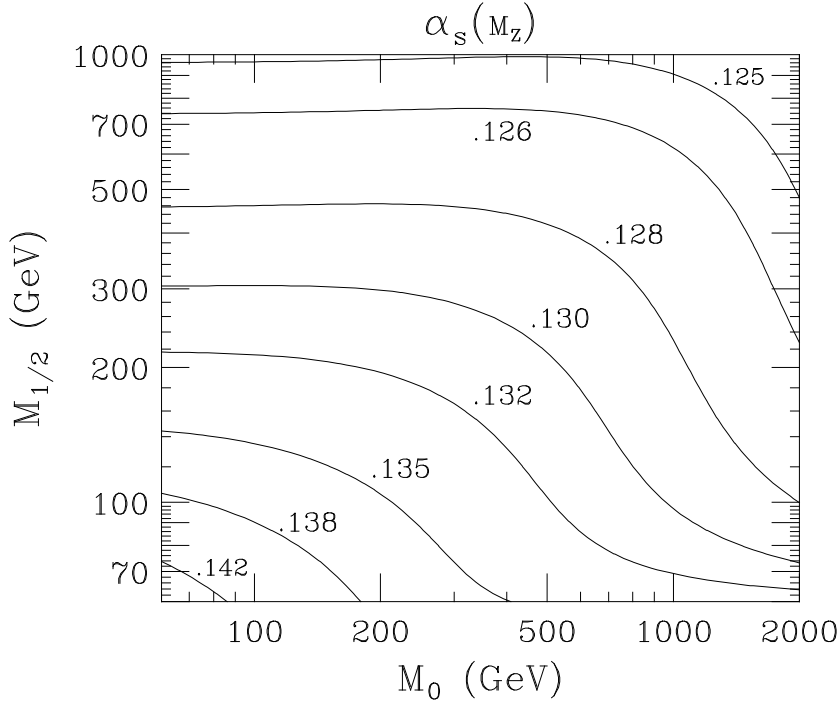}
\begin{center}
\parbox{5.5in}{
\caption[]{\small Contours of $\alpha_s(M_Z)$ in the $M_0$, $M_{1/2}$
plane with $m_t=170$ GeV, $\tan\beta=2$ and $A_0=0$. The region
$m_{\tilde q}<1$ TeV is bounded by $M_{1/2}\ \roughly{<}\ 400$ GeV
and $M_0<1$ TeV.}}
\end{center}
\end{figure}

The values of $\alpha_s(M_Z)$ quoted here are larger than in many previous
analyses for two reasons.  First, during the past few years the central value
of $s^2_{\rm SM}$, as determined from precision electroweak measurements, has
been decreasing.  (This change is correlated with the increasing best-fit
value of $m_t$.)  A smaller value of $s^2_{\rm SM}$ leads to an increase in
$\alpha_s(M_Z)$, as can be seen from eq.~(\ref{delta_alpha}).  Second, the
finite corrections decrease
$\hat s^2$, and therefore increase $\alpha_s(M_Z)$.  The finite corrections are
important in the region $M_{1/2}\,\roughly{<}\,200$ GeV where $\alpha_s(M_Z)$
is appreciably larger than in the leading logarithmic approximation,
as shown in Fig.~2.

\begin{figure}[tb]
\epsfysize=3in
\epsffile[65 500 680 730]{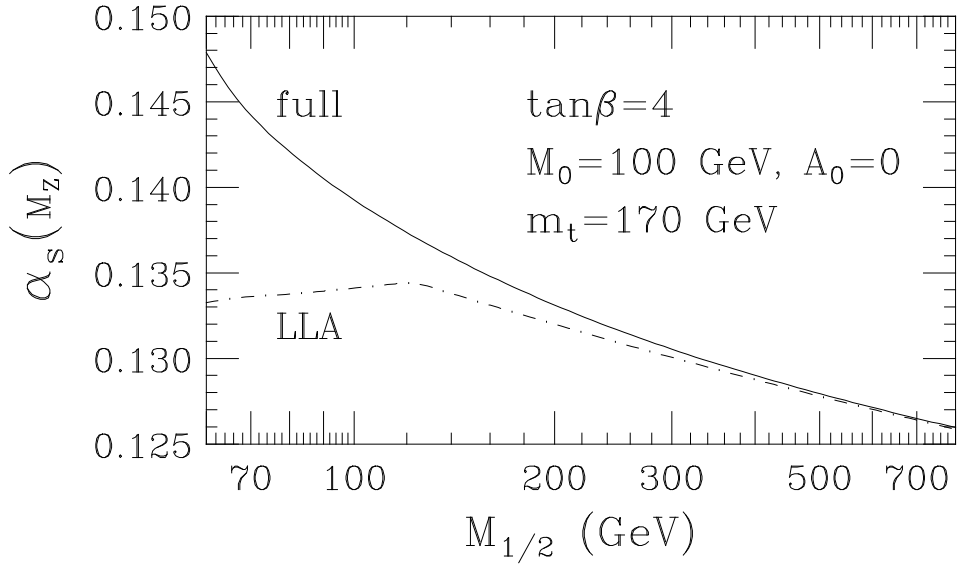}
\begin{center}
\parbox{5.5in}{
\caption[]{\small The \ms coupling $\alpha_s(M_Z)$ vs. $M_{1/2}$.
The curve labeled LLA shows the result if we include
only the logarithms of the supersymmetric masses (the leading logarithm
approximation), while the solid line corresponds to the full result
including all finite corrections.}}
\end{center}
\end{figure}

Of course, the value of $\alpha_s(M_Z)$ can be reduced by a unification-scale
threshold correction with $\epsilon_g  < 0$.   In the minimal SU(5) model
\cite{minimal su five}, the unification-scale gauge threshold correction is
\cite{min su five gauge corrs}
\begin{equation}
\epsilon'_g\ =\
{3g_{\rm GUT}^2\o40\pi^2}\,\log\left(M_{H_3}\o
M_{\rm GUT}\right)\ ,
\label{epsprime}
\end{equation}
where $M_{H_3}$ is the mass of the color-triplet Higgs fermion that mediates
nucleon decay.\footnote{The gauge threshold $\epsilon'_g$ also receives
contributions from higher-dimensional Planck-scale operators that we ignore.
They lead to a $\pm0.006$ uncertainty in $\alpha_s(M_Z)$
\cite{langacker and polonsky}.}
{}From this expression, we see that $\epsilon'_g < 0$ whenever
$M_{H_3} < M_{\rm GUT}$.  However, $M_{H_3}$ is bounded from below by proton
decay experiments, so $\epsilon'_g > 0$ throughout most of the parameter
space.\footnote{We use the formulae given in ref.~\cite{Hisano}, and
choose the conservative values $\beta=0.003$ GeV$^3$ and $|1+y^{tK}|=0.4$.}

In minimal SU(5) we find that the smallest possible values
for $\alpha_s(M_Z)$ are typically even larger than those in Fig.~1.
The only exception occurs for $M_0 \gg M_{1/2}$, where the proton
decay amplitude is suppressed. This determines the lower limit
$$\alpha_s(M_Z)\ >\ 0.123 \quad\quad\quad ({\rm Minimal\ SU(5)},
\ m_{\tilde q}\le1{\rm\ TeV},\ m_t=170 \rm\ GeV),$$
assuming $\hat\lambda_t(M_{\rm GUT}) \le 3$. In fact,
as long as $m_{\tilde q}\le1$ TeV,
$\alpha_s(M_Z)<0.127$ can only be obtained in the region
$M_0\simeq1$ TeV. For example, if $M_0\le500$ GeV,
$\alpha_s(M_Z)>0.128$.

The missing-doublet model is an alternative SU(5) theory in which
the heavy color-triplet Higgs particles are split naturally
from the light Higgs doublets \cite{missing doublet}.  In this
model the unification-scale gauge threshold
can be written as \cite{Yamada}
\begin{equation}
\epsilon_g^{\prime\prime}\ =
\ {3g_{\rm GUT}^2\o40\pi^2}\,\Biggl\{\log\left(M_{H_3}^{\rm eff}
\o M_{\rm GUT}\right) - {25\o2}\log5 + 15\log2\Biggr\}\ \simeq
\ \epsilon'_g - 4\%\ .
\label{mdmodel}
\end{equation}
Thus, for fixed $M_{H_3}$, the missing-doublet model has the same
threshold correction as the minimal SU(5) model, minus 4\%.  In
eq.~(\ref{mdmodel}), $M_{H_3}^{\rm eff}$ is the effective mass that enters
into the proton decay amplitude, so the bounds on $M_{H_3}$ in the minimal
SU(5) model also apply to $M_{H_3}^{\rm eff}$ in the missing-doublet
model.

The large negative correction in eq. (\ref{mdmodel}) gives rise to much smaller
values for $\alpha_s(M_Z)$.  This is illustrated in
Fig.~3, where we show the upper and lower bounds on $\epsilon_g$ in the minimal
and missing-doublet SU(5) models, together with the values of $\epsilon_g$
necessary to obtain $\alpha_s(M_Z) =0.117 \pm 0.01$.  For both SU(5) models
we bound $\epsilon_g$ from below by the limits on proton decay and from
above by the requirement $M_{{H_3}} < 10^{19}$ GeV.  Note that for the
missing-doublet model, the region of allowed $\epsilon_g$ nearly overlaps
the region
with $\alpha_s(M_Z) = 0.117 \pm 0.01$, regardless of the supersymmetric
particle masses.

\begin{figure}[tb]
\epsfysize=3.2in
\epsffile[55 250 680 530]{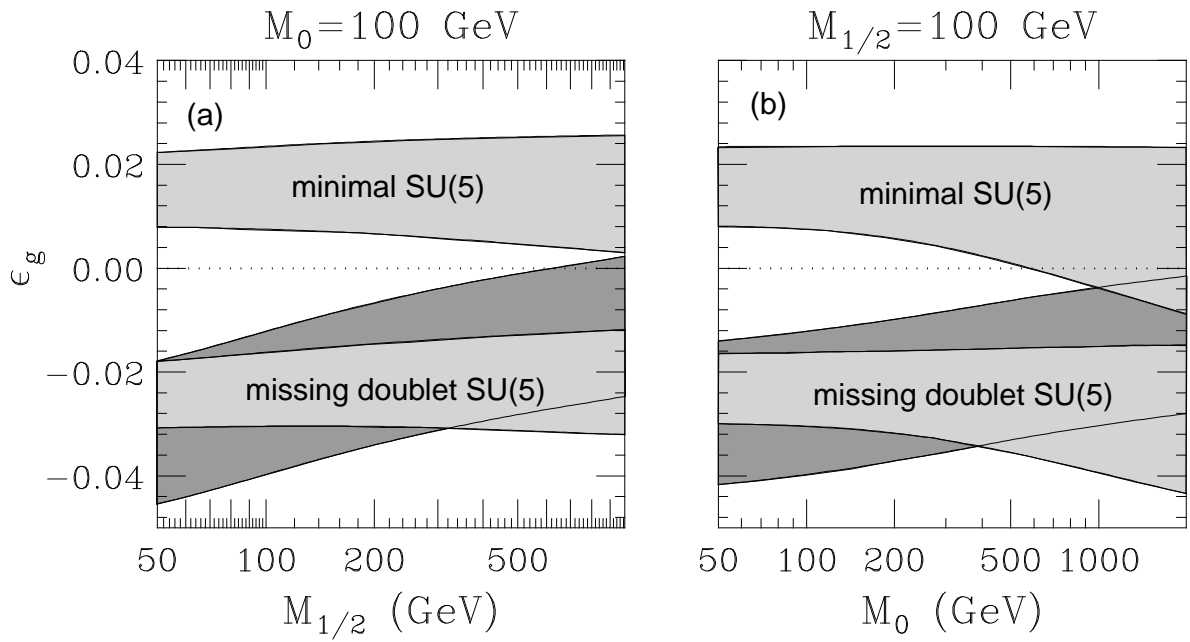}
\begin{center}
\parbox{5.5in}{
\caption[]{\small The light shaded regions indicate the allowed
values of the gauge coupling threshold correction $\epsilon_g$
in the minimal and missing-doublet SU(5) models.
The dark shaded region indicates the range of $\epsilon_g$
necessary to obtain $\alpha_s(M_Z) = 0.117\pm0.1$.}}
\end{center}
\end{figure}

\section{Yukawa Coupling Unification}

In typical SU(5) models the bottom and tau Yukawa couplings unify at the
scale $M_{\rm GUT}$.  In the remainder of this letter, we will investigate
bottom-tau unification to the same next-to-leading-order accuracy that we
used in our analysis of the gauge couplings.  Our analysis is based on
two-loop RGE's and on the complete one-loop expressions that relate the
\dr Yukawa couplings to the physical fermion masses \cite{BMPZ}.

Let us illustrate our procedure for the case of the tau Yukawa coupling.
We start by determining the \dr expectation value $\hat v$ from the $Z$-boson
mass,
\begin{equation}
{1\over4}\,(\hat g'^2 + \hat g^2)\hat v^2\ =\ M_Z^2 + \hat\Pi_Z(M_Z^2)\ .
\end{equation}
Then, given the pole mass $m_\tau = 1.777$ GeV \cite{PDG},
we find the tau Yukawa
coupling $\hat\lambda_\tau(M_Z)$ using the \dr relation
\begin{equation}
\hat \lambda_\tau(M_Z)\, \hat v \cos\beta/\sqrt{2}\ =\ m_\tau +
\hat\Sigma_\tau(m_\tau)\ ,
\end{equation}
where $\hat\Sigma_\tau(m_\tau) = \Sigma_1 + m_\tau \Sigma_\gamma$, and
the tau self-energy is $\Sigma_1 + \rlap/p \Sigma_\gamma +
\gamma_5\left(\Sigma_5 + \rlap/p \Sigma_{\gamma_5}\right)$. We follow a similar
procedure to find the top-quark Yukawa coupling $\hat\lambda_t(M_Z).$

Once we have the top and tau Yukawa couplings at the scale $M_Z$, we
evolve them to the scale $M_{\rm GUT}$ using the two-loop supersymmetric
RGE's \cite{two loop}.  At $M_{\rm GUT}$ we fix the \dr bottom-quark
Yukawa coupling $\hat\lambda_b(M_{\rm GUT})$ through the unification
condition
\begin{equation}
\hat\lambda_b(M_{\rm GUT})\ =\ \hat\lambda_\tau(M_{\rm GUT})(1+\epsilon_b)\ ,
\label{eq:epsb}
\end{equation}
where $\epsilon_b$ parametrizes the unification-scale Yukawa threshold
correction.  We then run all the Yukawa couplings back to the $Z$ scale,
and iterate the procedure self-consistently to determine $\hat\lambda_b(M_Z)$.
Finally, we apply the weak-scale threshold corrections to find the pole
mass for the bottom quark.

For a top-quark pole mass $m_t = 170$ GeV, this procedure typically gives a
bottom-quark mass
outside the range of experiment (which we take to be 4.7 $< m_b <$ 5.2 GeV
\cite{PDG}).  From previous analyses we know that for $m_t < 200$ GeV
there are two regions
of $\tan\beta$ where bottom-tau unification might
occur: $\tan\beta\ \roughly{<}
\ 2$ and $\tan\beta\ \roughly{>}\ 40$.  Here we focus on the small $\tan\beta$
branch, where the top-quark Yukawa coupling at the unification scale is large,
$\hat \lambda_t(M_{\rm GUT})\ \roughly{>}\ 1$.

As a point of reference, we first present our results with no unification-scale
threshold corrections.  In Fig.~4 we show $m_b$ and $\alpha_s(M_Z)$ versus
$m_t$, for various values of $\tan\beta$, $M_0$, and $M_{1/2}$, with
$\hat\lambda_t(M_{\rm GUT}) = 3$.  From the figure we see that the bottom-quark
pole mass is generally too large, unless the squark masses are of order 1 TeV.

\begin{figure}[tb]
\epsfysize=3in
\epsffile[0 400 680 690]{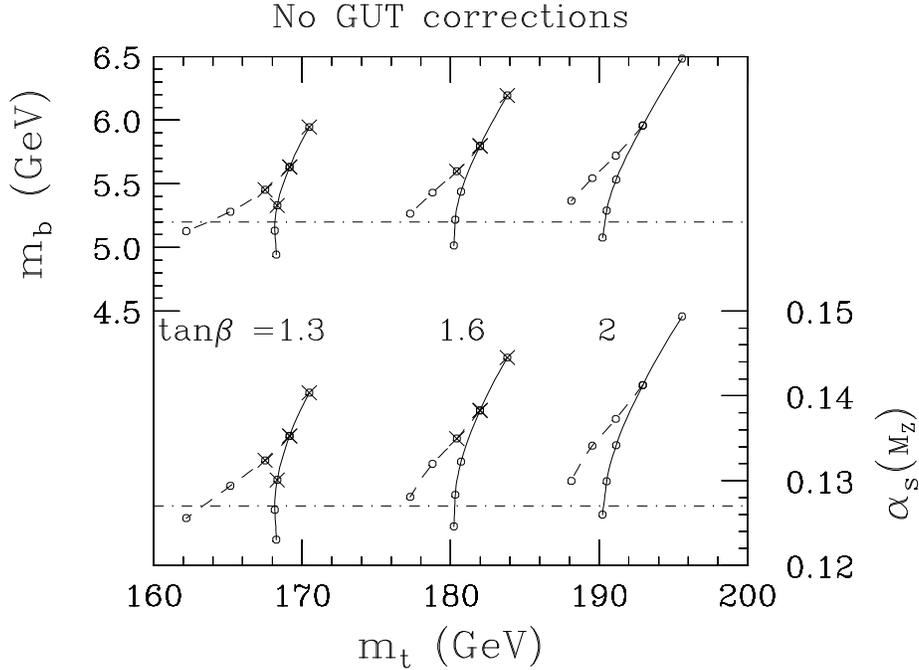}
\begin{center}
\parbox{5.5in}{
\caption[]{\small The bottom-quark mass and $\alpha_s(M_Z)$
vs. $m_t$ for the case of no unification-scale thresholds,
for various values of $\tan\beta$, with $A_0=0$ and
$\hat\lambda_t(M_{\rm GUT})=3$. The right (solid) leg in
each pair of lines corresponds
to $M_{1/2}$ varying from 60 to 1000 GeV, with $M_0$ fixed at 60 GeV.
The left (dashed) leg corresponds to $M_0$ varying from
60 to 1000 GeV, with $M_{1/2}=100$ GeV. On the solid lines the circles
mark, from top to bottom, $M_{1/2}=60$, 100, 200, 400, and 1000 GeV,
and on the dashed lines the circles mark $M_0=60$, 200, 400, and
1000 GeV. Note that the lowest point on each left leg and the
second-to-lowest point on each right leg corresponds to
$m_{\tilde q}\simeq1$ TeV. The horizontal dashed lines indicate
$m_b=5.2$ GeV and $\alpha_s(M_Z)=0.127$. The $\times$'s mark points
with one-loop Higgs mass $m_h<60$ GeV.}}
\end{center}
\end{figure}

The value $\hat\lambda_t(M_{\rm GUT}) = 3$ was chosen because it lies on the
edge of perturbation theory.  Smaller values of $\hat\lambda_t(M_{\rm GUT})$
give rise to larger values of $m_b$, so the curves in Fig.~4 can be interpreted
as lower limits on the bottom-quark mass.  If we require the squark masses to
be below 1 TeV, we obtain the lower bound
$$ m_b\ >\ 5.1 \ {\rm GeV}\quad\quad\quad ({\rm No\ unification\ thresholds},
\ m_{\tilde{q}}\le1{\rm\ TeV},\ m_t = 170 \rm\ GeV).$$

In the absence of unification-scale threshold corrections, we see that, in the
small $\tan\beta$ region, bottom-tau Yukawa unification favors large values of
$m_b$ and large values of the supersymmetric mass scale. The above bound
depends slightly on the top-quark mass, and sensitively on
$\hat\lambda_t(M_{\rm GUT})$. For $m_t=160$ GeV, the limit reduces by 0.1 GeV.
The bound increases by 0.3 GeV for $\hat\lambda_t(M_{\rm GUT})=2$
and by 0.8 GeV for $\hat\lambda_t(M_{\rm GUT})=1$. These variations
also apply for $\epsilon_b \ne 0$.

As with $\alpha_s(M_Z)$, the picture is altered by unification-scale
threshold corrections.  To understand their effects,
we first note the striking similarity
between the $m_b$ and $\alpha_s(M_Z)$ curves in Fig.~4, which implies that
the value of $m_b$ is tightly correlated with the value of $\alpha_s(M_Z)$.
This leads us to expect that the gauge threshold correction
will have an important effect on $m_b$.

In Fig.~5 we show the prediction for $m_b$ versus $m_t$, for fixed $\epsilon_b
= 0$, with $\hat\lambda_t (M_{\rm GUT}) = 2,3$ and $\epsilon_g$ chosen to give
a fixed value of $\alpha_s(M_Z)$.  We see that any model which gives
$\alpha_s(M_Z)
\simeq 0.12$ also predicts $m_b \simeq 5$ GeV, with $\epsilon_b = 0$.
Furthermore, any $-23\%<\epsilon_b<9\%$ gives an acceptable
value for $m_b$ in this case.  Alternatively, if $\alpha_s(M_Z) \simeq 0.13$,
a large and negative $\epsilon_b$ is required to achieve the central value
for $m_b$, namely $-33\%<\epsilon_b<-8\%$.
Figure 5 shows that these conclusions hold independently of the
top-quark mass and the supersymmetric mass scale.

\begin{figure}[tb]
\epsfysize=3in
\epsffile[0 408 700 708]{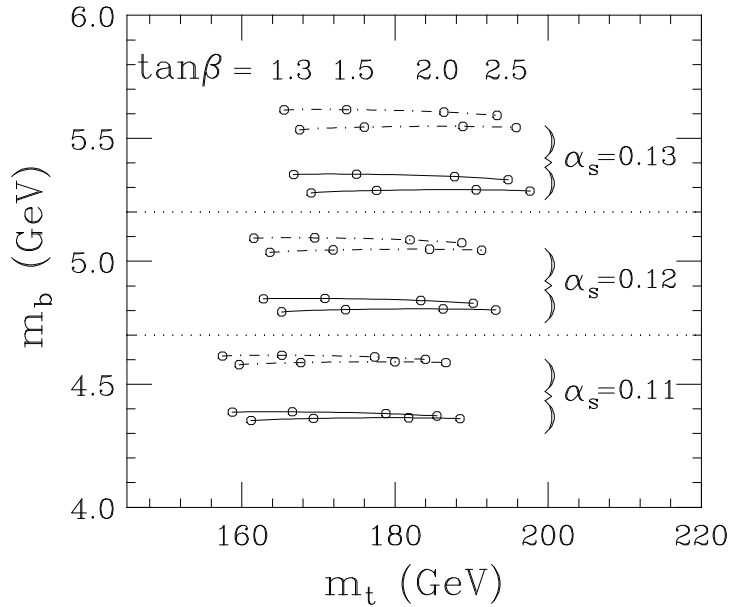}
\begin{center}
\parbox{5.5in}{
\caption[]{\small The bottom-quark mass vs. the top-quark mass
for fixed values of $\alpha_s(M_Z)$ and various $\tan\beta$.
The solid lines correspond to $\hat\lambda_t(M_{\rm GUT})=3$
while the dashed lines correspond to $\hat\lambda_t(M_{\rm GUT})=2$.
The upper line in each pair corresponds to a light
supersymmetric spectrum with $M_0=M_{1/2}=80$ GeV. The lower line
in each pair corresponds to a heavy spectrum,
$M_0=1000$ GeV, $M_{1/2}=500$ GeV.}}
\end{center}
\end{figure}

In Fig.~6 we plot the most favorable values of $\alpha_s(M_Z)$ and $m_b$ in
the minimal SU(5) model.  In this model the Yukawa threshold is given by
\cite{Wright}
\begin{equation}
\epsilon'_b\ =\ {1\over 16 \pi^2}\left[ \,4g_{\rm GUT}^2 \,\left(\log
\left({M_V\o M_{\rm GUT}}\right) - {1 \o 2} \right)\ -
\ \hat\lambda_t^2(M_{\rm GUT})\,\left(\log
\left({M_{H_3}\o M_{\rm GUT}}\right) - {1 \o 2} \right)\,\right]\ ,
\end{equation}
where $M_V$ is the mass of the superheavy SU(5) gauge bosons.
We define the most favorable value as follows.  We first
minimize the value of $\alpha_s(M_Z)$ by picking the smallest
possible $M_{H_3}$ consistent with nucleon decay.  We then minimize
$m_b$ by choosing the smallest $M_V$ allowed by the constraint that
the SU(5) model Yukawa couplings remain perturbative,
$M_V>\max(0.3 M_{H_3},0.5 M_{\rm GUT})$. From the figure we see
that this brings $m_b$ to an acceptable range for squark masses of
order 1 TeV.  Indeed, we find the limit\footnote{We could try to
decrease $m_b$  further by increasing $M_{H_3}$, which decreases
$\epsilon'_b$. However, the corresponding increase in $\alpha_s(M_Z)$
compensates for this so that the limit in eq.~(\ref{mb lim}) remains
unchanged.}
\begin{equation}
m_b\ >\ 5.1 \ {\rm GeV} \quad\quad\quad ({\rm Minimal\ SU(5)},
\ m_{\tilde{q}}\le1{\rm\ TeV},\ m_t=170 \rm\ GeV).\label{mb lim}
\end{equation}

\begin{figure}[tb]
\epsfysize=3in
\epsffile[0 408 700 708]{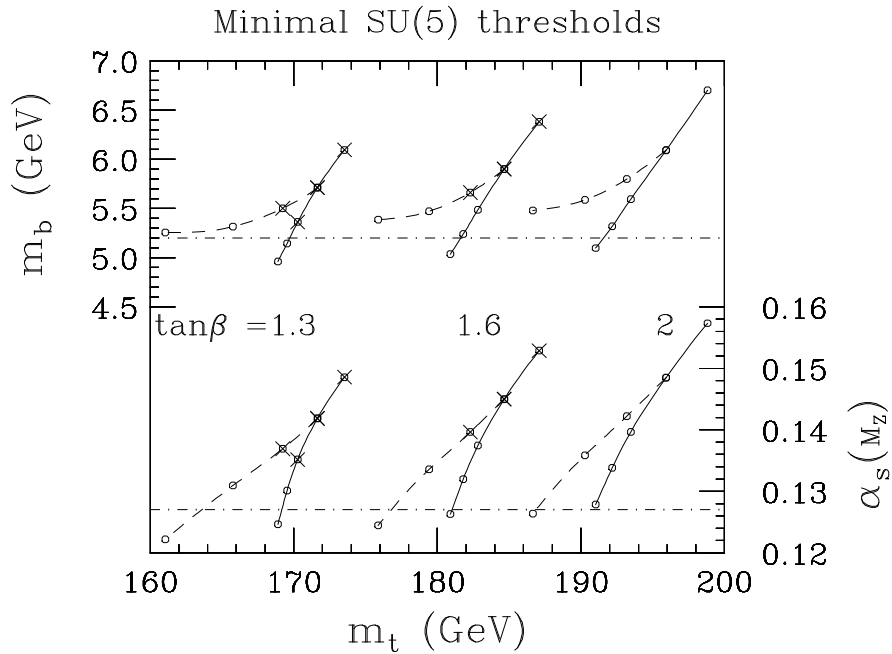}
\begin{center}
\parbox{5.5in}{
\caption[]{\small The same as Fig. 4, with the most favorable
minimal SU(5) threshold corrections, as defined in the text.}}
\end{center}
\end{figure}

In contrast to minimal SU(5), the missing-doublet model can readily
accommodate bottom-quark masses in the range $4.7 {\rm~GeV} < m_b
< 5.2 {\rm~GeV}$, with $\epsilon_b \simeq 0$.  For example, with $m_t=170$ GeV
and $M_0 = M_{1/2} = 100$ GeV, minimal SU(5) requires $\epsilon_b$ in the range
$-15$ to $-65\%$.  The missing-doublet model, however, requires $\epsilon_b$ in
the range $-30$ to $+20\%$.  Hence even for small
$\cal O$(100 GeV) supersymmetric masses the missing-doublet model naturally
accommodates both gauge and Yukawa coupling unification.

\section{Conclusion}

In this paper we have computed the complete one-loop weak-scale threshold
corrections in the minimal supersymmetric standard model.  We used them to
study gauge and Yukawa unification with and without unification-scale
threshold corrections.  In the absence of such corrections we find that
$\alpha_s(M_Z)$ and $m_b$ are large unless the squark masses are
larger than about 1 TeV.  Adding minimal SU(5) threshold
corrections, and requiring $m_{\tilde q}\le1$ TeV, we find
that $\alpha_s(M_Z)<0.127$ occurs only for $M_{1/2}\ll M_0$
and $M_0\simeq1$ TeV. Additionally, the condition $m_b<5.2$ GeV
is fulfilled only for $m_{\tilde q}\ \roughly{>}\ 1$ TeV.
In the missing-doublet model, however, the threshold
corrections permit acceptable values for $\alpha_s(M_Z)$
and $m_b$ for either a light or heavy supersymmetric spectrum.

\vskip .4in
\noindent {\Large\bf Acknowledgements}
\vskip .2in

We would like to thank R.~Zhang for collaboration during the
early stages of this work, and P.~Langacker and
S.~Pokorski for helpful discussions.  J.B. would like to thank Z.~P\l uciennik
for extensive discussions of preliminary results from ref.~\cite{Warsaw}
during his visit to Warsaw.  D.P. thanks H. Murayama for useful conversations
and B. Kniehl for helpful communications. This work was supported
by the U.S. National Science Foundation under grant NSF-PHY-9404057.

\end{document}